\documentclass[12pt]{JHEP3}
\usepackage{epsf,amsfonts,amssymb,epsfig,amsmath}

\renewcommand{\text}[1]{#1}

\newcommand{\be}{\begin{equation}}
\newcommand{\ee}{\end{equation}}
\newcommand{\ben}{\begin{displaymath}}
\newcommand{\een}{\end{displaymath}}
\newcommand{\bea}{\begin{eqnarray}}
\newcommand{\eea}{\end{eqnarray}}
\newcommand{\bean}{\begin{eqnarray*}}
\newcommand{\eean}{\end{eqnarray*}}
\newcommand{\nn}{\nonumber \\}
\newcommand{\ba}{\begin{array}}
\newcommand{\ea}{\end{array}}
\newcommand{\bi}{\begin{itemize}}
\newcommand{\ei}{\end{itemize}}

\newcommand{\reef}[1]{(\ref{#1})}

\newcommand{\call}{\mbox{${\cal L}$}}

\newcommand{\bbR}{{\mathbb{R}}}

\newcommand{\bbI}{{\mathbb{I}}}

\title{Constraining Maximally Supersymmetric Membrane Actions}

\author{Jerome P. Gauntlett \\
Theoretical Physics Group, Blackett Laboratory, \\
        Imperial College, London SW7 2AZ, U.K. \\
The Institute for Mathematical Sciences, \\
        Imperial College, London SW7 2PE, U.K. \\
        E-mail: \email{j.gauntlett@imperial.ac.uk}}
   
\author{Jan B. Gutowski \\
    DAMTP, Centre for Mathematical Sciences
University of Cambridge, \\
Wilberforce Road, Cambridge, CB3 0WA, UK   \\
E-mail: \email{J.B.Gutowski@damtp.cam.ac.uk}}

\abstract{We study the recent construction of maximally supersymmetric
field theory Lagrangians in three spacetime dimensions that are based on algebras
with a triple product. Assuming that the algebra has a positive definite metric compatible with the
triple product, we prove that the only non-trivial examples are either the well known case
based on a four dimensional algebra or direct sums thereof.}


\begin{document}

\setcounter{equation}{0}


\section{Introduction}

A better understanding of the three-dimensional superconformal field theory that
arises on multiple membranes in flat space is an important outstanding issue in M-theory.
Building on earlier work \cite{Schwarz:2004yj,Basu:2004ed}, an interesting
Lagrangian description of a maximally
supersymmetric conformal field theory in three dimensions was constructed
in \cite{Bagger:2006sk,Gustavsson:2007vu,Bagger:2007jr}
which has been further studied in
\cite{Bagger:2007vi} - \cite{Hosomichi:2008qk}.
The construction relies on an algebra with a skew triple product whose
structure constants $f^{\mu_1\mu_2\mu_3}{}_\nu= f^{[\mu_1\mu_2\mu_3]}{}_\nu$ satisfy
\be\label{one}
f^{\mu_1\mu_2\mu_3}{}_\nu f^{\mu_4\mu_5\nu}{}_{\mu_6}=3f^{\mu_4\mu_5[\mu_1}{}_\nu f^{\mu_2\mu_3]\nu}{}_{\mu_6}
\ee
or equivalently
\be\label{quadeq}
f^{[\mu_1\mu_2\mu_3}{}_\nu f^{\mu_4]\mu_5\nu}{}_{\mu_6}=0\ .
\ee
The construction of the Lagrangian requires a compatible metric
and, after raising an index on $f$ using this metric, $f$ is totally antisymmetric
$f^{\mu_1\mu_2\mu_3\mu_4}=f^{[\mu_1\mu_2\mu_3\mu_4]}$.
Since the metric appears in the kinetic terms
of the Lagrangian, it is natural to demand that the metric is positive definite. 
In this case, after a suitable change of basis, we can assume
that the metric is simply $\delta_{\mu\nu}$.
The basic non-trivial solution \cite{Bagger:2007jr} corresponds to a four dimensional algebra
with $f^{\mu_1\mu_2\mu_3\mu_4}=\epsilon^{\mu_1\mu_2\mu_3\mu_4}$.
One can also consider direct sums of this basic example, but this simply leads to
three-dimensional supersymmetric
field theories which are non-interacting copies of the basic example.

We started this work by trying to construct additional solutions to \reef{quadeq}
with totally antisymmetric $f$.
However, as also noticed by others, obvious generalisations fail and simple computer searches
are fruitless. It has also been shown \cite{FigueroaO'Farrill:2002xg}
that in up to seven dimensions, a 4-form whose components
satisfy ({\ref{quadeq}}) must be proportional to
$dx^{1234}$ (in some appropriately chosen co-ordinates),
and in eight dimensions, the solution is
a linear combination $dx^{1234}$ and $dx^{5678}$.

Here we will prove the general result, that all solutions of \reef{quadeq}, in any dimension,
can be written as a linear combination 4-forms, each of which is the wedge product
of four 1-forms, which are all mutually orthogonal. This then proves conjectures made in \cite{FigueroaO'Farrill:2002xg} and
\cite{Ho:2008bn}.

{\bf Note added:}  Concurrent with the  posting of this work to the ArXive,
a proof of this result also appeared in \cite{Papadopoulos:2008sk}. 
After this paper was submitted for publication, we became aware of
\cite{nagy2007}, which claims the same result using a different approach.

\section{Analysis}
We are interested in solutions to \reef{quadeq} for totally anti-symmetric and real $f$ with indices raised and lowered using
the metric $\delta_{\mu\nu}$. Let us assume that we have a $D+1$ dimensional algebra and write the indices
as $\mu = (q, D+1)$ where $q=1, \dots , D$. We can write
\be
f = dx^{D+1} \wedge \psi + \phi
\ee
where $\psi$ is a 3-form on $\bbR^D$, and $\phi$ is a 4-form on
$\bbR^D$. We can demand that $\psi\ne 0$ (otherwise we end up in $D$ dimensions).
The constraint ({\ref{quadeq}}) is equivalent to
\be
\label{eq1}
\phi^{[q_1 q_2 q_3}{}_m \phi^{q_4] q_5 q_6m}
+ \psi^{[q_1 q_2 q_3} \psi^{q_4] q_5 q_6}=0
\ee
\be
\label{eq2}
\phi^{[q_1 q_2 q_3}{}_m \psi^{q_4] q_5m} =0
\ee
\be
\label{eq3}
\phi^{q_1 q_2 q_3}{}_m \psi^{q_4 q_5m}
-3 \psi^{[q_1 q_2}{}_ m \phi^{q_3] q_4 q_5m}=0
\ee
\be
\label{jacob}
\psi^{[q_1 q_2}{}_m \psi^{q_3] q_4m} =0
\ee
where indices on $\psi$, $\phi$ are raised/lowered with $\delta_{mn}$.
Observe that ({\ref{jacob}}) is the Jacobi identity.
This identity implies that $\psi_{mn}{}^p$ are the structure constants of
a Lie algebra $\call$. The Killing form of this Lie algebra has components
\be
\kappa_{mn} = \psi_{m \ell}{}^p \psi_{n p}{}^\ell\ .
\ee
As $\psi$ is totally antisymmetric, note that $\kappa$ is negative semi-definite.
There are two possibilities:
$\kappa$ is non-degenerate and $\call$ is semi-simple or $\kappa$ is degenerate.

Suppose that $\call$ is semi-simple. By making a $SO(D)$ rotation, one can diagonalize
the Killing form and set
\be
\kappa_{mn} = -\lambda_n \delta_{mn}
\ee
(no sum over $n$), and $\lambda_n >0$ for all $n$.

On the other hand if $\kappa$ is degenerate, then $\call= u(1)^p \oplus \call'$ where $p>0$ and $\call$ is semi-simple.
To see this we first note that $X^m \kappa_{mn}=0$ for some non-zero vector $X^n$.
Then it follows that
\be
X^m X^n \psi_{m p q} \psi_n{}^{p q} =0
\ee
which implies that $X^n \psi_{npq}=0$. Without loss of generality, one can make an
$SO(D)$ rotation so that the only non-vanishing component of $X^n$ is $X^1$ and
then $\psi_{1mn}=0$ for all $m,n$, and $\kappa_{1m}=0$ for all $m$. By repeating this process
in the directions $2, \dots, D$ one finds after a finite number of steps,
either that  $\call= u(1)^p \oplus \call'$ where $p>0$ and $\call'$ is semi-simple, or $\psi=0$ which we
have assumed not to be the case.

We will analyse the two cases in turn,
but we first establish some useful identities arising from ({\ref{eq2}})-({\ref{jacob}})
that are valid in both cases.
We define $h=-\kappa$ i.e.
\be
h_{mn} = \psi_{mab} \psi_n{}^{ab}\ .
\ee
First contract ({\ref{eq2}}) with $\psi_{q_4 q_5 \ell}$ so that one obtains
\be
\label{compact1}
\phi^{q_1 q_2 q_3 m} h_{m \ell} - \phi^{q_4 q_2 q_3 m} \psi^{q_5 q_1}{}_m \psi_{q_5 q_4 \ell}
-  \phi^{q_1 q_4 q_3 m} \psi^{q_5 q_2}{}_m \psi_{q_5 q_4 \ell}
-  \phi^{q_1 q_2 q_4 m} \psi^{q_5 q_3}{}_m \psi_{q_5 q_4 \ell} =0\ .
\ee
However, note that the Jacobi identity implies that
\be
\label{jacobcon1}
\phi^{q_4 q_2 q_3 m} \psi^{q_5 q_1}{}_m \psi_{q_5 q_4 \ell} = {1 \over 2} \phi^{q_2 q_3 m n} \psi^{r q_1}{}_\ell
\psi_{r m n}\ .
\ee
Using this identity one can rewrite ({\ref{compact1}}) as
\be
\label{compact2}
\phi^{q_1 q_2 q_3 m} h_{m \ell}-{1 \over 2} \phi^{q_2 q_3 m n} \psi^{r q_1}{}_\ell
\psi_{r m n}
-{1 \over 2} \phi^{q_3 q_1 m n} \psi^{r q_2}{}_\ell
\psi_{r m n}
-{1 \over 2} \phi^{q_1 q_2 m n} \psi^{r q_3}{}_\ell
\psi_{r m n} =0\ .
\ee
Also, contracting ({\ref{eq2}}) with $\delta_{q_3 q_5}$ gives
\be
\label{compact3}
\phi^{q_1 q_2 m n} \psi^{q_4}{}_{m n} + \phi^{q_2 q_4 m n} \psi^{q_1}{}_{m n}
+ \phi^{q_4 q_1 m n} \psi^{q_2}{}_{m n} =0\ .
\ee
Next, contract ({\ref{eq3}}) with $\psi_{q_1 q_2 \ell}$ to obtain
\be
\label{compact4}
- \phi^{q_3 q_4 q_5 m} h_{m \ell} + \phi^{q_1 q_2 q_3 m} \psi_{q_1 q_2 \ell} \psi^{q_4 q_5}{}_m
-2 \phi^{q_2 q_4 q_5 m} \psi^{q_3 q_1}{}_m \psi_{q_1 q_2 \ell}=0\ .
\ee
This can be rewritten (using ({\ref{jacobcon1}}) to simplify the last term) as
\be
\label{compact5}
- \phi^{q_1 q_2 q_3 m} h_{m \ell} + \phi^{m n q_1 r} \psi_{m n \ell} \psi^{q_2 q_3 r}
+ \phi^{q_2 q_3 m n} \psi^{r q_1}{}_\ell \psi_{r m n} =0\ .
\ee
On contracting this expression with $\delta_{q_1 q_3}$, the first and the third term vanish
(the third term vanishes as a consequence of the Jacobi identity), and we find
\be
\label{compact6}
\phi^{n_1 n_2 m_1 m_2} \psi_{n_1 n_2 \ell} \psi_{m_1 m_2 r} =0\ .
\ee
Next, contract ({\ref{compact5}}) with $\psi_{q_2 q_3 s}$. The last term vanishes
as a consequence of ({\ref{compact6}}), and we obtain
\be
\label{compact7}
- \phi^{m n q r} h_{r \ell} \psi_{m n s} + \phi^{m n q r} h_{r s} \psi_{m n \ell} =0\ .
\ee

\subsection{Solutions when $\call$ is semi-simple}
We now assume that $\call$ is semi-simple. As we have already observed,
we can make a rotation and work in a basis for which
\be
h_{mn} = \lambda_n \delta_{mn}
\ee
(no sum over n), with $\lambda_n>0$ for all $n$.

Then ({\ref{compact7}}) implies
\be
\label{compact8}
- \phi^{m n q}{}_\ell \lambda_\ell \psi_{m n s} + \phi^{m n q}{}_s \lambda_s \psi_{m n \ell} =0
\ee
with no sum over $\ell$ or $s$.
On substituting this expression back into ({\ref{compact3}}) we obtain
\be
\label{compact9}
(\lambda_{q_4} - \lambda_{q_1}-\lambda_{q_2}) \phi^{q_1 q_2 m n} \psi^{q_4}{}_{mn}=0
\ee
(no sum on $q_1,q_2,q_4$).
Hence $\phi^{q_1 q_2 m n} \psi^{q_4}{}_{mn}=0$, or
$\lambda_{q_4} - \lambda_{q_1}-\lambda_{q_2}=0$
for some choice of $q_1, q_2, q_4$.
Now, it is not possible to have
$\lambda_{q_4} - \lambda_{q_1}-\lambda_{q_2}=\lambda_{q_1} - \lambda_{q_2}-\lambda_{q_4}
= \lambda_{q_2} - \lambda_{q_1}-\lambda_{q_4}=0$ simultaneously.
Hence, at least one of $\phi^{q_1 q_2 m n} \psi^{q_4}{}_{mn}$,
$\phi^{q_1 q_4 m n} \psi^{q_2}{}_{mn}$, $\phi^{q_2 q_4 m n} \psi^{q_1}{}_{mn}$
must vanish. However, ({\ref{compact8}}) then implies that {\it all} these terms vanish.
Hence we conclude that
\be
\label{compact10}
\phi^{q_1 q_2 m n} \psi^{q_4}{}_{mn}=0
\ee
for all $q_1, q_2, q_4$.
Finally, on substituting ({\ref{compact10}}) back into ({\ref{compact2}}), the last
three terms are constrained to vanish, hence
\be
\phi_{q_1 q_2 q_3 q_4}=0\ .
\ee

Now consider ({\ref{eq1}}). This implies that
\be
\psi^{[q_1 q_2 q_3} \psi^{q_4] q_5 q_6} =0
\ee
which implies (see e.g. \cite{FigueroaO'Farrill:2002xg})
that $\psi$ is simple i.e. it can be written as the wedge product of three one forms.
Hence one can chose a basis for which
\be
\psi= \lambda dx^1 \wedge dx^2 \wedge dx^3\ .
\ee
Furthermore, as $\call$ is compact, this implies that $\call$ must be 3-dimensional
i.e. $\call= su(2)$. We have thus recovered the
basic four-dimensional case with
$f^{\mu_1\mu_2\mu_3\mu_4}=\epsilon^{\mu_1\mu_2\mu_3\mu_4}$.

\subsection{Solutions when $\call$ is not semi-simple}

Set $\call = u(1)^p \oplus \call'$ where $p>0$ and $\call'$ is semi-simple.
It will be useful to split the indices $m$ into ``semi-simple" directions ${\hat{m}}$
and ``$u(1)$" directions $A$, so $m=({\hat{m}}, A )$.
Note that $\psi_{A m n}=0$ for all $m, n$, and $h_{A m}=0$ for all $m$,
but $h_{\hat{m} \hat{n}} = \lambda_{\hat{n}}\delta_{\hat{m} \hat{n}}$ (no sum on $\hat{n}$).
Recall the identity ({\ref{compact2}}).
Setting $q_1=A$, $q_2=B$, $q_3=C$ one finds
\be
\label{noncompact1}
\phi_{ABC {\hat{m}}} =0\ .
\ee
Also, setting $q_1=A$, $q_2=B$, $q_3={\hat{m}}$ one finds
\be
\label{noncompact2}
\phi^{AB \hat{m}\hat{s}}h_{\hat{s}\hat{\ell}} - \frac{1}{2} \phi^{AB \hat{p} \hat{q}} \psi^{\hat{s} \hat{m}}{}_{\hat{\ell}}
\psi_{\hat{s} \hat{p} \hat{q}} =0\ .
\ee
However, ({\ref{compact3}}) implies that
\be
\label{noncompact3}
\phi^{AB \hat{p} \hat{q}} \psi_{\hat{s} \hat{p} \hat{q}} =0
\ee
and so on substituting this back into ({\ref{noncompact2}}) one finds
\be
\label{noncompact4}
\phi_{AB \hat{m} \hat{n}} =0\ .
\ee
Returning to the general conditions \reef{eq2}, \reef{eq3} and \reef{jacob} with all free indices hatted,
we can follow the same steps in the last subsection to conclude that
\be
\phi_{\hat{m}\hat{n}\hat{p}\hat{q}}=0\ .
\ee
Thus the only non-zero components of $\phi$ are of the form $\phi^{A\hat{q_1}\hat{q_2}\hat{q_3}}$ and $\phi^{ABCD}$.

Considering other indices in \reef{eq2}, \reef{eq3} and \reef{jacob} we conclude that
\be
\label{reducjord}
\psi^{[\hat{q_1}\hat{q_2} }{}_{\hat{m}}\psi^{\hat{q_3}]\hat{q_4}\hat{m}}=0
\ee
\be
\label{mixed1}
\phi^{A\hat{q_1}\hat{q_2}}{}_{\hat{m}}\psi^{\hat{q_3}\hat{q_4}\hat{m}}
=\phi^{A\hat{q_3}\hat{q_4}}{}_{\hat{m}}\psi^{\hat{q_1}\hat{q_2}\hat{m}}
\ee
\be
\label{mixed2}
\phi^{A\hat{q_1}[\hat{q_2}}{}_{\hat{m}}\psi^{\hat{q_3}\hat{q_4}]\hat{m}}=0\ .
\ee
From \reef{eq1} we also get
\be
\label{mixed3}
\phi^{[A_1A_2A_3}{}_B\phi^{A_4]A_5A_6B}=0
\ee
\be
\label{mixed4}
\phi^{\hat{q_1}\hat{q_2}\hat{q_3}}{}_B\phi^{A_1A_2A_3B}=0
\ee
\be
\label{mixed5}
\phi^{A[\hat{q_1}\hat{q_2}}{}_{\hat{m}}\phi^{\hat{q_3}]\hat{q_4}B\hat{m}}=0
\ee
\be
\label{mixed6}
\phi^{\hat{q_1}\hat{q_2}}{}_{\hat{m}[A_1}\phi_{A_2]}{}^{\hat{q_3}\hat{q_4}\hat{m}}=0
\ee
\be
\label{mixed7}
\psi^{[\hat{q_1}\hat{q_2}\hat{q_3}}\psi^{\hat{q_4}]\hat{q_5}\hat{q_6}}
+\phi^{[\hat{q_1}\hat{q_2}\hat{q_3}}{}_A\phi^{\hat{q_4}]\hat{q_5}\hat{q_6}A}=0\ .
\ee

To proceed with the analysis, it is convenient to define the matrices $T^A$ by
\be
(T^A)_{\hat{m}}{}^{\hat{n}} = \phi^{A \hat{q}_1 \hat{q}_2 \hat{n}} \psi_{\hat{q}_1 \hat{q}_2 \hat{m}}\ .
\ee
On contracting ({\ref{mixed1}}) with $\delta_{\hat{q}_2 \hat{q}_4}$, we observe that
$T^A$ are all symmetric matrices. Furthermore, on contracting
({\ref{mixed6}}) with  $\delta_{\hat{q}_2 \hat{q}_4}$ and making use of
 ({\ref{mixed1}}), it is straightforward to show that the matrices $T^A$ commute with each other.
 Also, ({\ref{mixed1}}) implies that the $T^A$ commute with $h$.

Next, note that the Jacobi identity ({\ref{reducjord}}) implies that
\be
(T^A)_{\hat{m} \hat{\ell}}  \psi^{\hat{\ell}}{}_{\hat{p} \hat{q}}= \phi^{A \hat{s} \hat{t}}{}_{\hat{m}} \psi_{\hat{s} \hat{t}
\hat{\ell}} \psi^{\hat{\ell}}{}_{\hat{p} \hat{q}}
= -2 \phi^{A \hat{s} \hat{t}}{}_{\hat{m}} \psi_{\hat{s} \hat{p} \hat{\ell}}
\psi^{\hat{\ell}}{}_{\hat{q} \hat{t}}
\ee
However, now using \reef{mixed1} and then the Jacobi identity again,
we get
\be
-2 \phi^{A \hat{s} \hat{t}}{}_{\hat{m}} \psi_{\hat{s} \hat{p} \hat{\ell}}
\psi^{\hat{\ell}}{}_{\hat{q} \hat{t}} =
-2\phi^{A\hat{s}}{}_{\hat{p}\hat{l}}\psi_{\hat{s}}{}^{\hat{t}}{}_{\hat{m}}\psi^{\hat{l}}{}_{\hat{q}\hat{t}}
=
-\phi^{A\hat{s}\hat{l}}{}_{\hat{p}}\psi_{\hat{s}\hat{l}\hat{t}}{}\psi^{\hat{t}}{}_{\hat{m}\hat{q}}\ .
\ee
Thus
\be
\label{mixed8}
(T^A)_{\hat{m} \hat{\ell}}  \psi^{\hat{\ell}}{}_{\hat{p} \hat{q}} =
 -(T^A)_{\hat{p} \hat{\ell}} \psi^{\hat{\ell}}{}_{\hat{m} \hat{q}}\ .
 \ee

Next, decompose semi-simple $\call' = \call_1 \oplus \dots \oplus \call_m$
where $\call_i$ are simple ideals such that $\call_i \perp \call_j$ (with respect to $h$),
and $[ \call_i , \call_j ]=0$
if $i \neq j$, and the restriction of the adjoint rep. to $\call_i$ is irreducible;
furthermore, $h|_{\call_i} = 2 \mu_i^2 \bbI$ for $\mu_i \neq 0$.
Contract ({\ref{mixed1}}) with $\psi_{\hat{q}_3 \hat{q}_4 \hat{\ell}}$ to obtain
\be
\phi^{A \hat{q}_1 \hat{q}_2 \hat{m}} h_{\hat{m} \hat{\ell}}
= \phi^{A \hat{q}_3 \hat{q}_4 \hat{m}} \psi^{\hat{q}_1 \hat{q}_2}{}_{\hat{m}}
\psi_{\hat{q}_3 \hat{q}_4 \hat{\ell}}\ .
\ee
Suppose that the indices ${\hat{q}}_1, {\hat{q}}_2$ lie in two different ideals $\call_i, \call_j$
for $i \neq j$. Then the RHS of the above expression vanishes, hence
for these indices, $\phi_{A \hat{q}_1 \hat{q}_2 \hat{m}}=0$, for all $\hat{m}$.
 Similarly, for these indices
$
(T^A)_{\hat{q}_1}{}^{\hat{q}_2} = \phi^{A \hat{r} \hat{\ell} \hat{q}_2} \psi_{\hat{r} \hat{\ell} \hat{q}_1} =0
$.

Consider $T^A_i$, the restriction of $T^A$ to $\call_i$.
Then  ({\ref{mixed8}}) implies that $T^A_i$ commutes with the restriction of the
adjoint rep. to $\call_i$. However, as this restriction of the adjoint rep. is irreducible,
it follows by Schur's Lemma that
\be\label{schurs}
T^A_i = \lambda^A_i \bbI\ .
\ee
As the $T^A$ all commute, this can be achieved for all $T^A$.

Next, consider ({\ref{mixed7}}) with all ${\hat{q}}$ indices restricted to $\call_i$.
Contracting this expression with $\psi_{\hat{q}_1 \hat{q}_2 \hat{q}_3}
\psi_{\hat{q}_5 \hat{q}_6 \hat{m}}$ gives
\be
\bigg( \sum_A (\lambda^A_i)^2 + 4 (\mu_i)^4 \bigg) \bigg( {\rm dim \ } \call_i -3 \bigg) \delta_{\hat{m}}^{\hat{q}_4}
=0
\ee
which implies that ${\rm dim \  } \call_i =3$ for all $i$, so $\call_i = su(2)$.
It follows that
\be
\psi = \sum_i \mu_i \theta_i
\ee
with $\mu_i\ne 0$, where
\be
\theta_i = dy_i^1 \wedge dy_i^2 \wedge dy_i^3
\ee
If the ${\hat{q}}$ indices are restricted to $\call_i$, since dim $\call_i=3$, $\phi_{A\hat{q_1}\hat{q_2}\hat{q_3}}$ must be proportional to
$\theta_i$. The proportionality constant can be fixed from \reef{schurs} and we
find
\be
\phi_{A \hat{q}_1 \hat{q}_2 \hat{q}_3} = {\lambda^A_i \over 2 \mu_i} (\theta_i)_{\hat{q}_1 \hat{q}_2
\hat{q}_3}\ .
\ee
It is convenient to re-define $\lambda^A_i = 2 \mu_i \chi^A_i$, so that
\be
\label{formeq}
f = dx^{d+1} \wedge \psi + \sum_{i,A} \chi^A_i dz^A \wedge \theta_i + \Phi
\ee
where $\Phi$ lies entirely in the $u(1)$ directions, whose directions we have denoted by $z^A$.
The remaining content of ({\ref{mixed7}}) is obtained by restricting
the indices $\hat{q}_1, \hat{q}_2, \hat{q}_3$ to $\call_i$,
and  $\hat{q}_4, \hat{q}_5, \hat{q}_6$ to $\call_j$
for $i \neq j$; we find
\be
\label{finalconstr}
\mu_i \mu_j + \sum_A \chi^A_i \chi^A_j   =0\ .
\ee
Note that the form $\Phi$ satisfies the quadratic constraint ({\ref{mixed3}}),
whereas ({\ref{mixed4}}) is equivalent to
\be
\label{finalorthog}
\chi^A_i \Phi_{A MNP}=0
\ee
for all $i$.

There are then two cases to consider. In the first case,
$\chi_i^A=0$ for all $A, i$. Then ({\ref{finalconstr}}) implies that
$\call'= su(2)$, and hence
\be
f = \mu_1 dx^{d+1} \wedge dy_1^1 \wedge dy_1^2 \wedge dy_1^3 + \Phi
\ee
where $\Phi$ has no components in the
$x^{d+1}, y_1^1, y_1^2, y_1^3$ directions.

In the second case, there exists some $A, i$ with $\chi^A_i \neq 0$.
Without loss of generality, take $i=1$. By making an
$SO(p)$ rotation entirely in the $u(1)$ directions, without loss of generality set
\be
\label{rotatu1}
\chi_1^1 = \tau,  \qquad \chi_1^A =0 \quad {\rm if \ A>1 }
\ee
where $\tau \neq 0$. Then, if $j \neq 1$, ({\ref{finalconstr}})
implies that
\be
\chi^1_j = -{\mu_1 \over \tau} \mu_j \ .
\ee
Substituting these constraints back into ({\ref{formeq}}), and rearranging the terms, one finds
\bea
f &=& (\mu_1 dx^{d+1} + \tau dz^1)\wedge \theta_1
+ \tau^{-1} (\tau dx^{d+1} - \mu_1 dz^1) \wedge \sum_{j>1} \mu_j \theta_j
\nn
&+& \sum_{j>1, A>1} \chi_j^A dz^A \wedge \theta_j  + \Phi\ .
\eea

Writing
\bea
f_1 &=&  (\mu_1 dx^{d+1} + \tau dz^1)\wedge \theta_1
\nn
{\tilde{f}} &=& \tau^{-1} (\tau dx^{d+1} - \mu_1 dz^1) \wedge \sum_{j>1} \mu_j \theta_j
+ \sum_{j>1, A>1} \chi_j^A dz^A \wedge \theta_j  + \Phi
\eea
we have found $f= f_1 +{\tilde{f}}$ where, as a consequence of
({\ref{finalorthog}}) and ({\ref{rotatu1}}), it follows that $\Phi$ has no components in the $z^1$ direction.

So, in both cases, we have the decomposition
\be
f= f_1 + {\tilde{f}}
\ee
where $f_1$ is a simple 4-form, and $f_1$, ${\tilde{f}}$ are totally orthogonal
i.e. $f_1^{\mu_1\mu_2\mu_3\nu}\tilde f^{\mu_4\mu_5\mu_6}{}_\nu=0$.

Having obtained this result, it is straightforward to prove
that if such an $f$ satisfies ({\ref{quadeq}}), then
\be
\label{decomp1}
f = \sum_{s=1}^N f_s
\ee
where $f_s$ are totally orthogonal simple 4-forms.
The proof proceeds by induction on the spacetime dimension
$D$ ($D \geq 4$). The result is clearly true for $D=4$. Suppose it is true for
$4 \leq D \leq d$. Suppose that $D=d+1$. Then by the previous reasoning, one
has the decomposition $f=f_1+ {\tilde{f}}$, where $f_1$ is a simple 4-form,
and $f_1, {\tilde{f}}$ are totally orthogonal. It follows that ${\tilde{f}}$ must
satisfy ({\ref{quadeq}}). Then either ${\tilde{f}}=0$ and we are done, or
${\tilde{f}}$ is a nonzero 4-form in dimension $d-3$, in which case it
follows that one can decompose ${\tilde{f}}$ into a finite sum of orthogonal
simple 4-forms, each of which is also orthogonal to $f_1$.

Hence we conclude that the decomposition ({\ref{decomp1}}) holds for all 4-forms
$f$ satisfying ({\ref{quadeq}}).

\section{Discussion}
Given the results presented here, the maximally supersymmetric
field theory Lagrangian based on the four-dimensional algebra with
$f^{\mu_1\mu_2\mu_3\mu_4}=\epsilon^{\mu_1\mu_2\mu_3\mu_4}$ is rather enigmatic.
If it is not to be an isolated curiosity, the assumptions going into the general
constructions of \cite{Bagger:2006sk,Gustavsson:2007vu,Bagger:2007jr} need to be relaxed.
One possibility is to relax the condition that the metric living
on the algebra is positive definite and some discussion recently appeared
in \cite{Ho:2008bn}. 
A different possibility is to not demand a Lagrangian description, but to work
instead at the level of the field equations and
this was recently discussed in \cite{Gran:2008vi}.
Another possibility, which also does not use totally antisymmetric structure constants, 
was considered in \cite{Morozov:2008cb}.

\acknowledgments{
We would like to thank Per Kraus for discussions.
JPG is supported by an EPSRC Senior Fellowship and a
Royal Society Wolfson Award.}

\end{document}